\documentclass[aps,prl,twocolumn,groupedaddress]{revtex4}

\usepackage{amsmath,bm}

\newcommand{\eeq}{\end{equation}}
\newcommand{\beq}{\begin{equation}}
\newcommand{\ba}{\begin{array}}
\newcommand{\ea}{\end{array}}
\newcommand{\bea}{\begin{eqnarray}}
\newcommand{\eea}{\end{eqnarray}}
\newcommand{\vev}[1]{\langle #1\rangle}

\begin{document}

\preprint{}

\title{E{\" o}tv{\" o}s bounds on couplings of fundamental parameters to gravity}

\author{Thomas Dent}

\affiliation{Institute for Theoretical Physics, University of Heidelberg, 
Philosophenweg 16, D-69120 Heidelberg, Germany}

\date{\today}

\begin{abstract} \noindent
The possible dependence of fundamental couplings and mass ratios on the gravitational potential has been bounded by comparing atomic clock frequencies over Earth's elliptical orbit. Here we evaluate bounds on such dependence from E{\" o}tv{\" o}s-type experiments that test the Weak Equivalence Principle, including previously neglected contributions from nuclear binding energy. We find that variations of fundamental parameters correlated with the gravitational potential are limited at $10^{-8}$--$10^{-9}$, an improvement of 2--3 orders of magnitude over atomic clock bounds.
\end{abstract}


\maketitle

\noindent
Both the Standard Model of particle physics and the theory of General Relativity are constructed on the assumption of Local Position Invariance (LPI): that locally measurable properties of matter do not vary over space and time. LPI forms part of Einstein's equivalence principle. While most tests of LPI concerning particle coupling strengths and masses have yielded results consistent with zero variation \cite{Uzan:2002vq}, there is a significant indication of cosmological variation in the fine structure constant $\alpha$ deduced from astrophysical absorption spectra \cite{Murphy:2003mi,Murphy:2003hw}. Such claims are controversial \cite{Srianand:2007zz,Molaro:2007kp,Murphy:2007qs,Murphy:2006vs} and await improvements in data and analysis. The proton-electron mass ratio $\mu\equiv m_p/m_e$ has also been probed: analysis of molecular H$_2$ spectra indicates a moderately significant variation \cite{Reinhold:2006zn,Ivanchik:2005ws} at high redshift, while recently a strong bound on variation at lower redshift has been derived from the NH$_3$ inversion spectrum \cite{Flambaum:2007fa}. 

It is also of interest to probe possible variations at the present time within the Solar System. Indeed any theory with an underlying Lorentz invariance, in which time variation may occur, should also allow spatial variation. 
Recently bounds have been set \cite{Blatt:2008su,Fortier:2007jf} on variations of fundamental parameters correlated with the gravitational potential $U$, by comparing atomic clock frequencies over several months while the Earth moves through the Sun's gravitational field \cite{Flambaum:2007ar,Shaw:2007ju}. For independently varying parameters $G_i$ we may define couplings $k_i$ via
\beq
	\Delta \ln G_i = k_i \Delta U
\eeq
for small changes in $U$; the annual variation due to Earth's orbital eccentricity is $\Delta U\simeq 3\times 10^{-10}$. 
The couplings of $\alpha$, $\mu$ and the light quark mass $m_q/\Lambda_c\equiv (m_u+m_d)/2\Lambda_c$, where $m_u$ and $m_d$ are the up and down quark current masses and $\Lambda_c$ is the invariant strong interaction scale of QCD \cite{Flambaum:2006ip}, were found to be consistent with zero \cite{Blatt:2008su}, with 
uncertainties
\beq
	\{\sigma(k_\alpha),\sigma(k_\mu),\sigma(k_q)\}= \{3.1,17,27\}\times 10^{-6}.
\eeq 
Now as pointed out in \cite{PeeblesDicke}, 
a spatial gradient of couplings or masses will lead to anomalous accelerations of test bodies, due to the dependence of their mass-energy on the varying parameters. 
With the gravitational acceleration $\vec{g}$ given by $-\vec{\nabla}U$, we have
$	\vec{\nabla}\ln G_i = - k_i \vec{g}.$
Rather than moving on a geodesic with acceleration $\vec{g}$, a freely-falling body of mass $M(G_i(\vec{x}))$ will experience additional acceleration $\vec{a}=-(\vec{\nabla} M)/M$ \cite{Brans:1961sx}, as if moving in a potential $V(\vec{x})= M(G_i(\vec{x}))$.

\paragraph{Units and definitions}
For a test body with a given space-time dependent mass $M_b({\bf x})\equiv M_0(1+\Delta M({\bf x})/M_0)$, we may perform a 
redefinition of units \cite{DickeMach} such that the value of $M_b$ is everywhere constant. The Newton ``constant'' $G_{\rm N}$ is then in general space-time dependent, and 
the body moves along geodesics of the rescaled metric $g'_{\mu\nu}=(M_0({\bf x})/M_0)^2g_{\mu\nu}$ \cite{Brans:1961sx}. 
In the Newtonian limit we have $g'_{00}\simeq -1-2U(\vec{x})-2\Delta M(\vec{x})/M_0$, and the previous result for $\vec{a}$ is recovered. For two test bodies of different $\bf{x}$-dependence such a redefinition {\em cannot}\/ be performed and the differential acceleration $\vec{a}_b-\vec{a}_c$, violating the Weak Equivalence Principle (WEP) or universality of free fall, is a physical signal probing spatial gradients of $G_i$.

For convenience we will express masses in Planck units where $G_{\rm N}$ is constant. 
The acceleration is then \cite{Damour:1994zq}
\beq
	\vec{a}_b = -\sum_i \frac{\partial \ln M_b}{\partial \ln G_i} \vec{\nabla}\ln G_i
	= \sum_i \frac{\partial \ln M_b}{\partial \ln G_i} k_i \vec{g},
\eeq
and the differential acceleration is
\beq
	\eta_{b-c}\equiv \frac{|\vec{a}_b-\vec{a}_{c}|}{|\vec{g}|}
	= \sum_i \frac{\partial \ln (M_b/M_c)}{\partial \ln G_i} k_i 
	\equiv \sum_i \lambda^{b-c}_i k_i.
\eeq
which defines the sensitivity coefficients $\lambda^{b-c}_i$. Each experimental limit on $\eta$ then bounds some linear combination of the couplings $k_i$. To bound $n$ independent couplings $k_i$, at least $n$ independent pairs of test bodies with different derivatives $\partial \ln M^{b,c}/\partial \ln G_i$ are needed. 

We may also change basis from one set of parameters $G_i$ to another $G'_k$, such that \footnote{Unless otherwise stated, the Einstein summation convention will apply to parameter indices $i$, $j$, {\it etc.}.}
\beq \label{basischange}
	k_i = F_{ik}k'_k,\qquad F_{ik} = \frac{\partial\ln G_i}{\partial \ln G'_k}
\eeq
for couplings $k'_k$ in the new basis. 

It is also essential to consider a {\em complete set}\/ of parameters $G_i$ which account for all non-negligible sources of variation in $M_b/M_c$, otherwise measurements of $\eta$ may be incorrectly interpreted. Couplings to baryon number $A$, lepton number $L=Z$ 
and electromagnetic self-energy $E_{em}\simeq a_C Z(Z-1)/A^{1/3}$ have commonly been considered \cite{Damour:1996xt}. However, there are other contributions to $M_b$ which depend on different, linearly independent functions of $A$ and $Z$. 
These arise from the strong nuclear binding energy, which for large $A$ is well approximated by the liquid drop model:
\beq \label{Bnuc}
	\frac{B_{nuc}}{A} = a_V - \frac{a_S}{A^{1/3}} - a_A\frac{(A-2Z)^2}{A^{2}} 
	+ a_P\frac{(-1)^A\!+\!(-1)^Z}{A^{3/2}} + \cdots,
\eeq
where the coefficients of volume, surface, asymmetry, and pairing energy respectively are $\{a_V, a_S, a_A, a_P\}\simeq\{15.7,17.8,23.7,11.2 \}\,$MeV \cite{Rohlf:1994wy}.
Such terms could {\em a priori}\/ couple to $U$ and induce differential acceleration. 

\paragraph{Statistical treatment}
We may construct \cite{Damour:1996xt,DamourBlaser} a likelihood function
\beq \label{chisq}
	\chi^2(k_i) = \sum_{\{b-c\}} \left(\frac{\eta_{b-c}-\lambda^{b-c}_i k_i}{\sigma_{b-c}}\right)^2, 
\eeq
where the sum over $\{b-c\}$ counts independent experimental results 
with central values $\eta_{b-c}$ and $1\sigma$ uncertainties $\sigma_{b-c}$. The likelihood ${\mathcal L}$ is proportional to $e^{-\chi^2/2}$. A ``metric'' $g_{ij}$ may be defined by the coefficient of the quadratic term $k_ik_j$ in Eq.~(\ref{chisq}):
\beq
	g_{ij} = \sum_{\{b-c\}} \sigma_{b-c}^{-2} \lambda^{b-c}_i \lambda^{b-c}_j.
\eeq 
The best-fit point with minimal $\chi^2=\chi^2_{min}$ is then
\beq
	k_{i,min} = (g^{-1})_{ij} \sum_{\{b-c\}} \sigma_{b-c}^{-2} \eta_{b-c} \lambda^{b-c}_j 
\eeq 
and the excess of $\chi^2$ above this minimum is given by
\beq \label{dchisq}
	\Delta \chi^2 \equiv \chi^2 - \chi^2_{min} =  g_{ij} \Delta k_i \Delta k_j,
\eeq
where $\Delta k_i \equiv k_i-k_{i,min}$. Error ellipsoids around the central values are then surfaces of constant $\Delta \chi^2$. The eigenvectors of $\bm g$ give the linear combinations of $k_i$ pointing along the principal axes of the ellipsoids.

We may project the likelihood onto each $k_i$ separately \cite{Fischer:2004jt,Blatt:2008su} by integrating over $k_j$ ($j\neq i$), resulting in a Gaussian distribution with variance 
\beq \label{projections}
	\sigma_i^2 = \frac{M_{ii}}{\det {\bm g}},
\eeq
where $M_{ii}$ is the minor of the element $g_{ii}$. 
On changing the basis of parameters via Eq.~(\ref{basischange}), the metric becomes
\beq \label{gtransf}
	g'_{km} = g_{ij}F_{ik}F_{jm} = ({\bm F}^T{\bm g}{\bm F})_{km}.
\eeq

\paragraph{Nuclear parameters and couplings}
To calculate $\lambda^{b-c}_i$ we identify the parameters controlling possible variations in mass ratios. We first consider ``nuclear parameters'' which characterize physics at low energy. 

The dimensionless parameters $X_I$ and their couplings to $U$ are: the fine structure constant $\alpha$ (coupling $k_\alpha$); the electron mass $m_e/m_N$ ($k_e$); the nucleon mass difference $\delta_N/m_N$ ($k_{\delta N}$); and the nuclear surface tension $a_S/m_N$ ($k_{aS}$), where $\delta_N\equiv m_n-m_p$ and $m_N\equiv (m_n+m_p)/2$. Further couplings may be defined for other terms in Eq.~(\ref{Bnuc}). 
The normalized mass per nucleon of a body is then
\begin{multline}
	\frac{M}{Am_N}= 1 - \left(f_p-\frac{1}{2}\right)\frac{\delta_N}{m_N} 
	+ f_p \frac{m_e}{m_N} + \\ 
	+ \frac{Z(Z-1)}{A^{4/3}}\frac{a_C}{m_N} 
	- \frac{a_V}{m_N} + A^{-1/3}\frac{a_S}{m_N} + \cdots\;
\end{multline}
where $f_p \equiv Z/A$, the coefficient $a_C\simeq 0.71\,$MeV varies proportional to $\alpha$, and the remaining terms are subleading in $A$. Expanding in small quantities we find
\begin{multline} \label{DeltalnMbMc}
	\Delta \ln \frac{M_b}{M_c} = \Delta \ln \frac{M_bA_c}{M_cA_b} 
	\\ \simeq \left ( -\frac{\delta_N}{m_N}
	\Delta \ln\frac{\delta_N}{m_N} 
	+ \frac{m_e}{m_N}
	\Delta \ln \frac{m_e}{m_N} \right)\, \hat{\Delta}_{b-c}f_p \\
	+ \frac{a_C}{m_N}
	\Delta \ln \alpha\, \hat{\Delta}_{b-c} \frac{Z(Z-1)}{A^{4/3}}
	+ \frac{a_S}{m_N} 
	\Delta \ln \frac{a_S}{m_N}\, \hat{\Delta}_{b-c} A^{-1/3}, 
\end{multline}
where $\hat{\Delta}_{b-c}$ denotes the difference of a quantity between two test body materials \footnote{For composite test masses, the differences of the averaged values $\vev{f_p}$, $\vev{Z(Z-1)A^{-4/3}}$ and $\vev{A^{-1/3}}$ apply.}. 
The sensitivity coefficients $\lambda^{b-c}_I$ may then be read off.
We cannot distinguish between couplings to $\delta_N/m_N$ and $m_e/m_N$, since their coefficients are both proportional to $\hat{\Delta}_{b-c} f_p$. Thus we define $Q_n\equiv \delta_N-m_e$ (the kinetic energy in neutron decay) with a coupling $k_{Qn}\equiv \Delta\ln (Q_n/m_N)/\Delta U$.
Its sensitivity coefficient is then $\lambda^{b-c}_{Qn} = -({Q_n}/{m_N}) \hat{\Delta}_{b-c}f_p.$

\paragraph{Results}
We summarize current E{\" o}tv{\" o}s experiments, specifying their pairs of test materials.

{\bf Schlamminger {\em et al.}}\ \cite{Schlamminger:2007ht}: Be--Ti, $\eta = (0.3\pm 1.8)\times 10^{-13}$. 
Be--Al data were also taken (to be published).

{\bf Bae\ss ler {\em et al.}}\ \cite{Baessler:1999iv}: Fe--SiO$_2$, $\eta = (0.5\pm 9.4)\times 10^{-13}$. 
Test bodies were `Earth core' Fe--Cr alloy and `Moon/mantle' silica body with a small fraction of Mg.

{\bf Y.~Su {\em et al.}}\ \cite{Su:1994gu}: 
\begin{align}
	\mbox{Be--Al},\ \eta &= (-0.2 \pm 2.8)\times 10^{-12} \nonumber \\
	\mbox{Be--Cu},\ \eta &= (-1.9 \pm 2.5)\times 10^{-12} \nonumber \\
	\mbox{Si/Al--Cu},\ \eta &= (5.1\pm 6.7)\times 10^{-12}\nonumber \\ 
 &\mbox{(Si dominant in Si/Al body)}. \nonumber
\end{align}

{\bf Braginsky and Panov} \cite{Braginsky72}: Pt--Al, $\eta = (-0.3\pm 0.4)\times 10^{-12}$. 
Here we rederived a $68\%$ confidence interval from the 7 quoted data points.
The meaning of the sign is ambiguous; we treat the result conservatively as a limit $|\eta| \leq 7\times 10^{-13}$ ($1\sigma$).

\begin{table}[b]
\caption{Sensitivity functions for test bodies. For SiO$_2$ we give a mass-weighted average.\label{bodiestable}}
\begin{ruledtabular}
\begin{tabular}{c|cccccc}
	Material & $A$ & $Z$ & $f_p-0.5$ & $\frac{Z(Z-1)}{A^{4/3}}$ & $A^{-1/3}$ 
	& $\frac{(A-2Z)^2}{A^2}$ \\
	\hline
	\hline
	Be & 9     & 4  & -0.0556 & 0.64 & 0.481 & 0.0123 \\
	Al & 27    & 13 & -0.0185 & 1.93 & 0.333 & 0.0014 \\
	Si & 28.1  & 14 & 0       & 2.14 & 0.329 & 0.0000 \\
	\hline
	SiO$_2$& 21.6 & 10.8 & 0  & 1.74 & 0.365 & 0.0000 \\
	\hline
	Ti & 47.9  & 22 & -0.042  & 2.65 & 0.275 & 0.0066 \\
	Fe & 56    & 26 & -0.0357 & 3.03 & 0.261 & 0.0051 \\
	Cu & 63.6  & 29 & -0.044  & 3.20 & 0.251 & 0.0078 \\
	Pt & 195.1 & 78 & -0.1    & 5.31 & 0.172 & 0.0402 \\
\end{tabular} 
\end{ruledtabular}
\end{table} 
The relevant functions of $A$ and $Z$ are listed in Table~\ref{bodiestable} \footnote{We do not consider Cr or Mg since these elements were subdominant in test mass composition and are sufficiently similar to Fe and Si, respectively.}.
The resulting sensitivity metric ${\bm g}$ in the basis $k_I=(k_{Q_n},k_\alpha,k_{aS})$ is
\beq \label{gIJ}
	{\bm g} = 
	\begin{pmatrix}
	0.00015 & -0.0020 & 0.0101  \\
	-0.0020 & 0.8696  & -2.030 \\
	0.0101  & -2.030  & 4.949 \\
	\end{pmatrix} \times 10^{20}.
\eeq
The eigenvalues are $\lambda_A=(7.0\times 10^{14},\, 3.2\times 10^{18},\, 5.8\times 10^{20})$, and we find that the linear combinations 
\beq
	\begin{pmatrix} \hat{k}_1 \\ \hat{k}_2 \\ \hat{k}_3 \end{pmatrix} 
	= \begin{pmatrix}
		0.998 & -0.058 & -0.026 \\
		0.064 & 0.923  & 0.381 \\
		0.002 & -0.382 & 0.924 \\
	\end{pmatrix} \begin{pmatrix} k_{Qn} \\ k_\alpha \\ k_{aS} \end{pmatrix}
\eeq
of couplings are bounded at the $10^{-7}$, $10^{-9}$ and $10^{-10}$ level respectively  \footnote{Very similar results follow from considering only the three most sensitive experiments (Be--Ti, Fe--SiO$_2$, Al--Pt).}.

Projecting the likelihood onto each direction $k_I$ via Eq.~(\ref{projections}), we obtain null bounds on three couplings with uncertainties
\beq 
	\{\sigma(k_{Qn}),\sigma(k_\alpha),\sigma(k_{aS})\} = 
	\{38, 2.3, 1.0\}\times 10^{-9},
\eeq
at least two orders of magnitude stronger than current atomic clock bounds, although we note that WEP experiments probe slightly different linear combinations of parameters from clock comparisons \cite{Damour:1997bf}. 

\paragraph{Including asymmetry energy}
In the above calculation we neglected the asymmetry term in Eq.~(\ref{Bnuc}) whose sensitivity coefficient is $\lambda^{b-c}_{aA} = \frac{a_A}{m_N} \hat{\Delta}_{b-c} ((A-2Z)^2/A^2)$.
If we include this contribution with an independent coupling $k_{aA}$, we find a 4-by-4 sensitivity metric $\bm g$ with eigenvalues $(3.4\times 10^{14}, 3.1\times 10^{15}, 5.6 \times 10^{18}, 5.8\times 10^{20})$, and the bounds projected onto each coupling separately are
\beq
	\{\sigma(k_{Qn}),\sigma(k_\alpha),\sigma(k_{aS}),\sigma(k_{aA})\} = 
	\{52,14,5.9,19\}\times 10^{-9}.
\eeq
The increase in uncertainty occurs because four experiments are now required to solve for the couplings $k_I$, but there are currently only three independent bounds with precision exceeding $10^{-12}$. 

Given the common origin of the surface and asymmetry terms in nuclear forces, we may assume that the coefficients $a_S/m_N$ and $a_A/m_N$ vary in the same way. Then we have $k_{aA}=k_{aS}\equiv k_{nuc}$ and we may rewrite with
\beq \label{lambdanuc}
	\lambda^{b-c}_{aS} k_{aS} + \lambda^{b-c}_{aA} k_{aA} = \lambda^{b-c}_{nuc} k_{nuc}
\eeq
where $\lambda_{nuc}= \lambda_{aS}+\lambda_{aA}$. We obtain a 3-by-3 metric in the basis $(k_{Qn},k_\alpha,k_{nuc})$ with small corrections compared to Eq.~(\ref{gIJ}). The eigenvalues are $(6.5\times 10^{14},5.6\times 10^{18},6.0 \times 10^{20})$ and the bounds projected onto each coupling separately are now 
\beq
	\{\sigma(k_{Qn}),\sigma(k_\alpha),\sigma(k_{nuc})\} = 
	\{39,1.7,0.8\}\times 10^{-9}.
\eeq
Hence with reasonable physical assumptions, bounds on $k_I$ are stable, or even improve, under increasingly precise approximations to nuclear binding energy.

\paragraph{Fundamental parameters and couplings}
We now consider how the ``nuclear parameters'' $X_I = (Q_n/m_N,\alpha,a_S/m_N)$ depend on parameters $G_k$ of the Standard Model of particle physics. 
These parameters are: the fine structure constant $\alpha$ (coupling $k'_\alpha$); the electron mass $m_e/\Lambda_c$ ($k'_e$); the light quark mass $m_q/\Lambda_c$ ($k'_{q}$); the up-down mass difference $\delta_q/\Lambda_c \equiv (m_d-m_u)/\Lambda_c$ ($k'_{\delta q}$); and the strange quark mass $m_s/\Lambda_c$, where we normalise to the QCD strong interaction scale. 

With three independent ``nuclear'' couplings $(k_{Qn},k_\alpha,k_{nuc})$, we can only bound the couplings of three independent combinations of fundamental parameters. The dependence of nucleon masses and nuclear forces on $m_s/\Lambda_c$ is subject to large uncertainties (possibly 100\%): a reliable calculation of the effects of varying $m_s/\Lambda_c$ is a formidable challenge. To proceed further, we must assume that $m_s/\Lambda_c$ does not couple significantly to $U$. 

The dependence of nucleon masses on $G_k$ was discussed in \cite{Dent:2006fn,Dent:2007zu} using results from chiral perturbation theory \cite{Gasser:1982ap,Borasoy:1996bx}.
We find, suppressing the term in $m_s/\Lambda_c$,
\beq
	\Delta \ln \frac{m_N}{\Lambda_c} \simeq 0.048\, \Delta \ln 
	\frac{m_q}{\Lambda_c} + (2.7\times 10^{-4})\, \Delta \ln \alpha +\cdots  
\eeq
For the isospin-violating mass difference $Q_n$ we derive
\begin{multline}
	\Delta \ln \frac{Q_n}{m_N} \simeq 2.6\Delta \ln\frac{\delta_q}{\Lambda_c} 
	-0.65\Delta \ln\frac{m_e}{\Lambda_c} \\ 
	-0.97 \Delta\ln \alpha 
	-0.048\Delta \ln \frac{m_q}{\Lambda_c},
\end{multline}
thus $k_{Qn}\simeq 2.6 k'_{\delta f} -0.97k'_\alpha -0.048k'_{q}$, where $k'_{\delta f} \equiv k'_{\delta q}-0.25k'_{e}$. Clearly also $k_\alpha=k'_\alpha$. 

\paragraph{Nuclear binding energy}
The dependence of strong nuclear binding energy on QCD parameters is subject to  significant uncertainties. For the deuteron a detailed effective field theory was applied \cite{Beane:2002xf,Epelbaum:2002gb} giving a dependence on the pion mass $m_\pi$ of
\beq
	\Delta \ln (B_{\rm D}/\Lambda_c) = r \Delta \ln (m_\pi/\Lambda_c) 
	\simeq \frac{r}{2} \Delta \ln (m_q/\Lambda_c),
\eeq
where $-10<r<-6$ and we use the leading order dependence $m_\pi\propto \sqrt{m_q}$. Taking this relation as a first approximation to the dependence of other nuclear binding energies \cite{Dent:2007zu} 
we estimated
$	\partial B_i/\partial m_\pi = f_i (A_i-1) r (B_{\rm D}/m_\pi)$,
with $f_i\sim 1$. This is consistent with estimates from realistic nuclear interaction models in small nuclei \cite{Flambaum:2007mj} if $r\simeq -7$. An effective theory treatment of the central force in larger nuclei yields a strong negative dependence of $a_A$ on $m_\pi/\Lambda_c$ \cite{Donoghue:2006du} which would only strengthen the resulting bounds. Thus 
\beq
	\Delta \ln \frac{B_i}{\Lambda_c} \sim \frac{-7 B_{\rm_D} (A_i-1)}{2 B_i}
	\Delta \ln \frac{m_q}{\Lambda_c} \simeq -0.9\Delta\ln \frac{m_q}{\Lambda_c},
\eeq
taking $B_i/(A_i-1)\simeq 8.5\,$MeV. If, as argued above, all coefficients in Eq.~(\ref{Bnuc}) react in a similar way to variation of $m_q/\Lambda_c$, $k_{nuc}$ is determined simply by the behaviour of $B_i/m_N$ and we have $k_{nuc}\simeq -0.9k'_{q}$. Thus the matrix $F_{Ik}$ giving the dependence of $X_I$ on $G_k$ is 
\beq
	{\bm F}=
	\begin{pmatrix}
	2.6 & -0.97 & -0.05 \\
	0   &  1    &  0    \\
	0   &  0    & -0.9  \\
	\end{pmatrix}
\eeq
where the columns correspond to variations of parameters ($\delta_f/\Lambda_c$, $\alpha$, $m_q/\Lambda_c$).

Transforming the sensitivity metric ${\bm g}$ (including the asymmetry energy contribution) via Eq.~(\ref{gtransf}) we obtain a metric ${\bm g'}$ for the couplings ($k'_{\delta f}$,$k'_{\alpha}$,$k'_q$). This yields null bounds with uncertainties
\beq
	\{\sigma(k'_{\delta f}),\sigma(k'_{\alpha}),\sigma(k'_q)\} = 
	\{14,1.7,0.9\}\times 10^{-9}. 
\eeq
This result is subject to theoretical uncertainty in the dependence of nuclear properties on quark masses, and is obtained with the assumption that variation of $m_s/\Lambda_c$ is negligible \footnote{Similar theoretical issues arise in interpreting atomic clock experiments, due to the uncertain dependence of nuclear magnetic moments on quark masses.}. However, it is clear that gravitational experiments testing WEP provide by far the most stringent bounds on variations of fundamental couplings correlated with the gravitational potential \cite{Nordtvedt:2002qe}. Significant improvements require space-based experiments \cite{Wolf:2007sh}, whether based on clocks or gravitation.

\begin{acknowledgments}
The author is supported by the {\em Impuls- and Vernetzungsfond der Helmholtz-Gesellschaft} and acknowledges enlightening discussions with Douglas Shaw, Stephan Schlamminger and Jan Schwindt, and correspondence with Thibault Damour.
\end{acknowledgments}


\end{document}